\documentclass[aps,prl,reprint,twocolumn,longbibliography,superscriptaddress, floatfix]{revtex4-2}
\usepackage{graphicx} 
\usepackage{dcolumn} 
\usepackage{bm}
\usepackage{amssymb} 
\usepackage[utf8]{inputenc}
\graphicspath{ {images/} }
\usepackage{mathtools}
\usepackage{physics}
\usepackage[english]{babel}
\usepackage{indentfirst}
\usepackage{graphicx}
\usepackage{amsfonts}
\usepackage{amssymb}
\usepackage{amsmath}
\usepackage{amsthm}
\usepackage{verbatim}
\usepackage{listings}
\usepackage{rotating}

\usepackage{float}
\usepackage[citecolor=blue, colorlinks=true, urlcolor=blue, linkcolor=blue]{hyperref}
\usepackage{booktabs}
\usepackage{csquotes}
\usepackage{color}

\usepackage{calc}
\usepackage{tabularx}
\usepackage{txfonts}
\usepackage{xcolor}
\usepackage{soul}

\newcommand{\fig}[1]{Fig.\thinspace{}\ref{#1}}

\newcommand{\fc}[1]{({#1})}
\newcommand{\figc}[2]{Fig.\thinspace{}\ref{#1}\thinspace{}\fc{#2}}

\usepackage{siunitx}
\usepackage{soul}
\usepackage[normalem]{ulem}
\graphicspath{{new/}}

\begin{document}

\newcommand{\TUM}{\affiliation{Technical University of Munich, TUM School of Natural Sciences, Physics Department, 85748 Garching, Germany}}
\newcommand{\MCQST}{\affiliation{Munich Center for Quantum Science and Technology (MCQST), Schellingstr. 4, 80799 M{\"u}nchen, Germany}}
\newcommand{\ETH}{\affiliation{Institute for Quantum Electronics, ETH Z\"urich, CH-8093 Z\"urich, Switzerland}}

\def\papertitle{{Tuning Unconventional Superconductivity with Feshbach Resonances in TMD Heterostructures}}
\def\papertitle{{Tuning Topological Superconductivity with Feshbach Resonances in Transition Metal Dichalcogenide Heterostructures}}

\def\papertitle{{Feshbach Tunable Topological Superconductivity in Transition Metal Dichalcogenide Heterostructures}}
\def\papertitle{{Realizing Topological Superconductivity in Tunable Bose-Fermi Mixtures with Transition Metal Dichalcogenide Heterostructures}}
\title{\papertitle}

\author{Caterina Zerba} \TUM \MCQST
\author{Clemens Kuhlenkamp} \TUM \MCQST \ETH
\author{Ata\c{c} Imamo\u{g}lu} \ETH
\author{Michael Knap} \TUM \MCQST

\date{\today}

\begin{abstract}
Heterostructures of two-dimensional transition metal dichalcogenides (TMDs) are emerging as a promising platform for investigating exotic correlated states of matter. Here, we propose to engineer Bose-Fermi mixtures in these systems by coupling inter-layer excitons to doped charges in a trilayer structure. Their interactions are determined by the inter-layer trion, whose spin-selective nature allows excitons to mediate an attractive interaction between charge carriers of only one spin species. Remarkably, we find that this causes the system to become unstable to topological p+ip superconductivity at low temperatures. We then demonstrate a general mechanism to develop and control this unconventional state by tuning the trion binding energy using a solid-state Feshbach resonance.

\end{abstract}

\maketitle

\begin{figure}[t]
\includegraphics[width=0.48\textwidth]{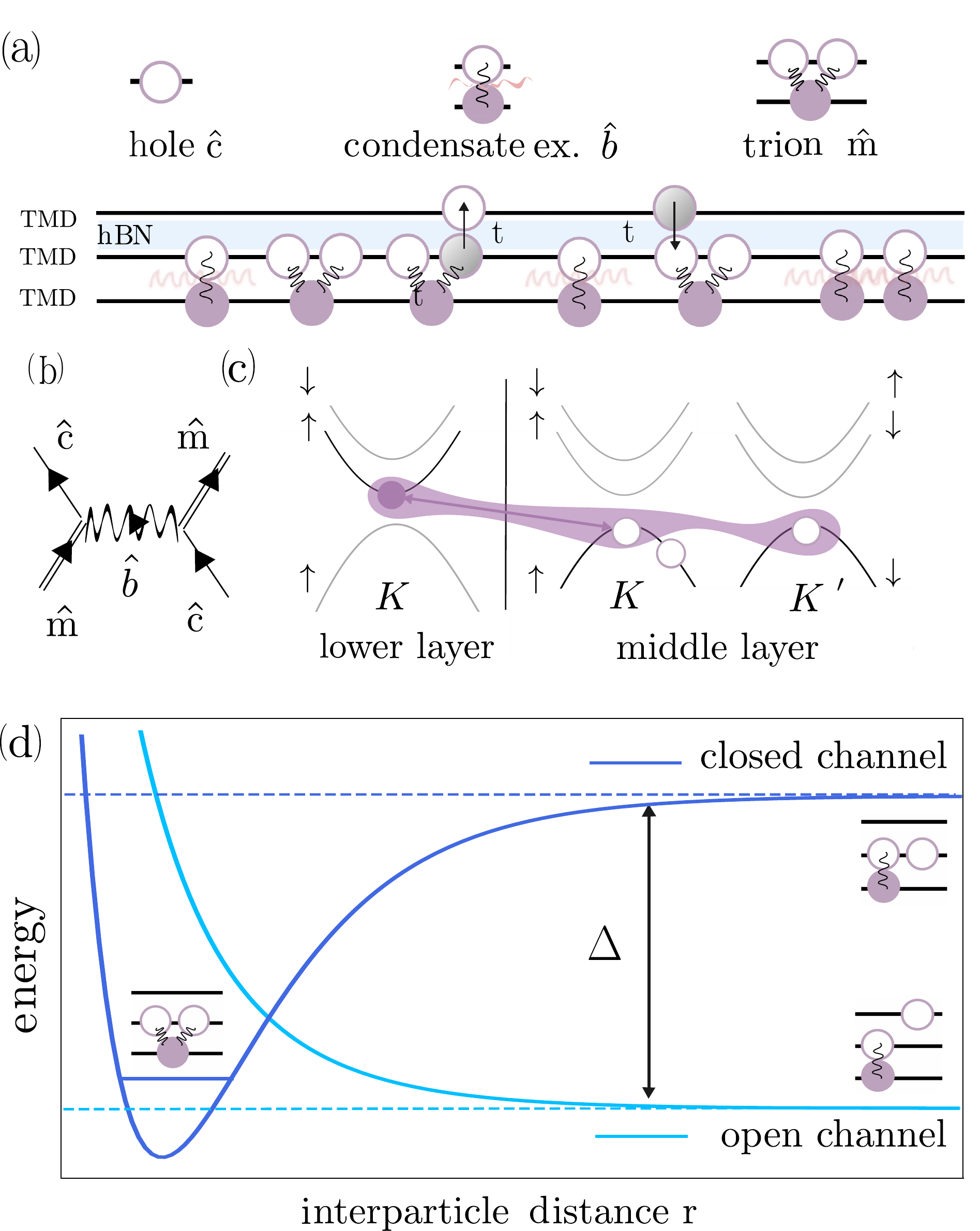}
\caption{\textbf{Tunable Bose-Fermi mixtures.} (a) We consider a TMD trilayer heterostructure. The relevant constituents are the charge carriers (holes), that coherently tunnel between the upper and the middle layer with strength $t$, an interlayer exciton condensate, and the trion bound state.  
(b) Effective attractive interactions between charge carriers are mediated by the virtual decay of a trion into a hole and a condensate excitation, which then scatters with another hole to create again a trion.
(c) The holes in the trion form a singlet. When the exciton condensate is created with a certain spin polarization, for example by circularly polarized light, attractive interactions are only mediated between holes of opposite spin. The exciton (electron-hole in say the $K$-valley) forms a trion with a $K'$-valley hole tunneled from the upper layer.}
(d) The superconducting state is controlled by a solid-state Feshbach resonance.  
Effective interactions are enhanced by tuning the threshold of the open channel close to the trion bound state in the closed channel by an external electric field $E_z \propto \Delta$. 
\label{fig:1}
\end{figure}
\textbf{Introduction.---}Heterostructures of two-dimensional semiconductors are particularly promising platforms to realize exotic phases of matter, such as correlated insulators, Mott-Wigner states, and superconductors~\cite{Cao2018, Cao2018Apr, Lu2019, Sharpe2019, Yankowitz2019, Shimazaki2020, Xu2020, Regan2020, Wang2020, Tang2020, Shimazaki2021, Li2021Wigner, Ma2021, Huang2021, Sun2021}. Due to their two-dimensional nature and high effective electron and hole masses, transition metal dichalcogenides (TMDs) in particular can host strongly correlated Bose-Fermi mixtures with fermions realized by doped charges and composite bosons by tightly bound excitons. Doped holes and inter-layer excitons interact by forming bound molecular states called trions. The bound-state formation in Bose-Fermi mixtures gives rise to a particularly rich phase diagram, potentially featuring among others coexisting liquids, unconventional superfluids, supersolids, and quantum phase transitions between them~\cite{Viverit2000, Powell05, Bchler2003, Enss2009, Fratini2010, Ludwig2011, Matuszewski2012, Cotle16, Kinnunen2018, Duda2023,Suzuki2008, Kalas2008}.  
In multi-layer semiconductor systems scattering between excitons and doped charges can be controlled and strongly enhanced when the energy of the trion is tuned by an external electric field, which is understood as a solid state Feshbach resonance~\cite{Clemens21,Schwartz2021ElectricallyTF}. Therefore, two-dimensional materials open new opportunities to realize and study Bose-Fermi mixtures in challenging low-temperature regimes. 

A natural route to search for unconventional superconducting states in TMDs~\cite{Hsu2017, Margalit2021, Lane2022, Scherer2022, Julku2022, Crepel2023, Li2021, Nayak2021}, such as  p-wave superconductivity which hosts exotic Majorana edge modes that are pertinent for topological quantum computing~\cite{Qi2011, Alicea_2012, Kallin_2016}, are thus provided by Bose-Fermi mixtures, as they allow for boson-mediated attractive interactions among charge carriers. 
However, a controlled realization of p-wave superconductivity remains challenging, as generically s-wave pairing is favored over higher angular momentum states. 

Here, we propose a general mechanism to develop and enhance unconventional p-wave superconductivity by tuning Bose-Fermi mixtures of interlayer excitons and holes in a TMD heterostructure. Superconductivity is achieved by hybridizing trions and holes, while the pairing is mediated by excitations of an exciton condensate; as illustrated in Fig.~\ref{fig:1}. Due to the spin-selective nature of the trion, only carriers of a single spin species interact attractively, which effectively restricts our model to spinless charge carriers. Then, the Pauli principle forces the Cooper pairs to carry odd angular momentum. Our analysis shows that the superconducting state is topological with p${}_x$+$i$p${}_y$ order. Moreover, the superconducting transition temperature is highly tunable and controllable by an external electric field via a Feshbach resonance.  
 
\textbf{Experimental setting.---}We consider a trilayer TMD heterostructure as shown in Fig.~\ref{fig:1} (a). Interlayer excitons can be optically injected in the bottom two layers at high densities ($\simeq 10^{12}$ cm$^{-2}$) due to their long lifetime~\cite{Jauregui2019Nov, Wilson2021}. They are strongly bound and essentially act as composite bosons. 
We assume that excitons are condensed~\cite{Wang2019Oct,Troue2023}. 
Fermions are introduced in the form of doped holes, which tunnel coherently between the top and middle layer. Hexagonal Boron Nitride (hBN) isolates the layers and, to a limited extent, controls the tunneling of charge carriers.
Inter-layer excitons and charge carriers can bind to a molecular trion state, which allows for controlling the boson-fermion scattering.
Tunable effective interactions between holes are then mediated by the exciton condensate; see~\figc{fig:1}{b}.  We specifically consider holes, as it has been found experimentally that the hole tunneling through hBN is much stronger than electron tunneling~\cite{Schwartz2021ElectricallyTF}.

We assume that the exciton condensate is optically injected in one valley by circularly polarized light, and thus through spin-momentum locking in one spin state, say the $K$-valley with spin-$\uparrow$; \figc{fig:1}{c}. Due to Pauli blocking, interlayer trion formation is only possible when the holes of the trion, that are both in the same layer, are in a spin singlet, as the triplet is generally unbound~\cite{Witham_2018, Tempelaar2019Jul, Dai_2023}. For small temperatures compared to the trion bound state energy, two charge carriers that interact with the same exciton to form a trion are thus necessarily in the same spin state. {This allows us to consider spinless fermions~\cite{Witham_2018, Tempelaar2019Jul, Dai_2023, Wang2019Oct, Rivera2015Feb}.}

The scattering process between holes and excitons can be controlled by a solid state Feshbach resonance~\cite{Clemens21, Schwartz2021ElectricallyTF}, where open and closed channels are tuned relatively to each other by an external electric field $E_z$; \figc{fig:1}{d}. Strong interactions are obtained when the open channel and the trion bound state are approximately in resonance. 

The system of holes $\hat c$, trions $\hat m$ (also referred to as molecules), and excitons $\hat x$ is described by the following effective Hamiltonian (see supplemental materials~\cite{supp}):
 \begin{equation}\label{eq1}
 \hat{H}= \hat H_\text{holes} + \hat H_\text{trions} +\hat H_\text{ex}+
 g\sum_{\mathbf{k},\mathbf{k}'} [\hat{m}^\dagger_{\mathbf{k}}\hat{c}_{\mathbf{k}'}\hat{x}_{\mathbf{k}-\mathbf{k}'} + \text{h.c.} ],
\end{equation}
where $\hat H_\text{holes} = \sum_{\mathbf{k}} \frac{k^2}{2 m} \hat{c}^\dagger_{\mathbf{k}} \hat{c}_{\mathbf{k}}$ is the kinetic energy of holes of mass $m$ and $\hat H_\text{trions}=\sum_{\mathbf{k}} \big(\frac{k^2}{2 M} +\Delta \big)\hat{m}^\dagger_{\mathbf{k}} \hat{m}_{\mathbf{k}}$ is the kinetic energy of trions of mass $M$ with $\Delta= e d E_z$ being the energy difference of a charge in the upper and the middle layer separated by distance $d$ that is tunable by the electric field $E_z$.  For holes and trions we consider an expansion near the minimum and maximum of the bands, at the $K$ and $K'$ points of the Brillouin zone, where their dispersion is parabolic to a good approximation. The interlayer excitons are governed by $\hat H_\text{ex} = \sum_{\mathbf{k}} \frac{k^2}{2 m_x} \hat{x}^\dagger_{\mathbf{k} }\hat{x}_{\mathbf{k}} + \frac{U}{2}\sum_{\mathbf{k},\mathbf{p},\mathbf{q}} \hat{x}_{\mathbf{k}+\mathbf{q}}^\dagger \hat{x}_{\mathbf{p}-\mathbf{q}}^\dagger \hat{x}_{\mathbf{k}}\hat{x}_{\mathbf{p}}$, where $m_x$ is the exciton mass and $U$ the strength of the exciton interactions that depends on the dielectric material between the two lowest layers and their distance~\cite{Jauregui2019Nov}. Since excitons are injected with low momenta and long-range electron-hole exchange is vanishingly small, due to spatial separation of the electron and the hole, a quadratic dispersion is justified. The last term in the Hamiltonian \eqref{eq1} describes the decay of the molecule into an exciton and a hole of strength $g$ that mediates the effective attractive interactions. Similar Hamiltonians have also been studied in the context of exciton-polariton mediated s-wave superconductivity~\cite{Laussy2010}.

We consider the regime in which the binding energy of the exciton is larger than the one of the trion. Then, the exciton's internal structure does not influence the properties of the trion. We are thus allowed to describe the excitons as a Bose gas~\cite{Amelio2022} and assume weak contact interactions. When the excitons condense~\cite{Fogler2014Jul, Gupta2020Jun}, we can expand $\hat{x}_{\mathbf{r}}=\sqrt{n_0} + \frac{1}{\sqrt{V}}\sum_{\mathbf{k} \not= 0 }e^{i\mathbf{k}\mathbf{r}} \hat{x}_{\mathbf{k}} $, where $n_0$ is the macroscopic condensate density. Performing a Bogoliubov analysis by expanding the purely excitonic part of the Hamiltonian, we obtain
\begin{align}
 \hat{H} &= \hat H_\text{holes} + \hat H_\text{trions} + \sum_{\mathbf{k}\not= 0} e_{x} (k) \hat{b}^\dagger_{\mathbf{k} }\hat{b}_{\mathbf{k}} \nonumber \\ &+g\sqrt{n_0} \sum_{\mathbf{k}} \big[\hat{m}^\dagger_{\mathbf{k}}\hat{c}_{\mathbf{k}} + \text{h.c.}\big]+g\sum_{\mathbf{k}\not=\mathbf{k}'} \big[\hat{m}^\dagger_{\mathbf{k}}\hat{c}_{\mathbf{k}-\mathbf{k}'}\hat{b}_{\mathbf{k'}} + \text{h.c.} \big], 
\end{align}
where $\hat{b}$ are the condensate excitations with energy $e_{x}(k)= \sqrt{\frac{k^2}{2 m_x}\big( \frac{k^2}{2 m_x}+ 2 U n_0 \big)} $. The condensate hybridizes the trion and charge bands with strength $g \sqrt{n_0}$. The three-body interaction contains sound excitations of the condensate $\hat b$, which are reminiscent of phonon mediated interactions. As the vertex $g$ already includes the tunneling process, the scattering effectively takes place in the same layer. Moreover, as the exciton’s Bohr radius is small, a recombination between the constituents can only occur locally when their wavefunctions have finite overlap. These considerations justify the assumption of contact interactions of strength $g$~\cite{Fey2020,Clemens21}.

The effective trion-exciton-hole coupling $g$ is fixed by the tunnel coupling $t$ of the electrons between the two layers as $g\simeq \sqrt{\frac{2\pi}{\mu_\text{red} |E_{T,0}|}} t$, where we estimate the bare trion  binding energy $E_{T,0} \simeq 6$ meV to be lower than the one reported in Ref.~\cite{Jauregui2019Nov} due to the additional hBN layer between the TMDs, and $\mu_\text{red}$ is the reduced mass of the hole and the exciton; see supplemental materials~\cite{supp}.  In our calculations, we assume the tunneling between the two layers to be  $t\simeq 0.9$ meV~\cite{Schwartz2021ElectricallyTF} . We typically consider exciton condensate densities of $n_0 \sim 10^{12}$ cm$^{-2}$ and estimate the interactions $U$ between the excitons using a plate capacitor formula as $U n_0 \simeq 10$ meV for $n_0 \sim 10^{12}$ cm$^{-2}$~\cite{Jauregui2019Nov}. For these densitiy regimes,  excitons are expected to be tightly bound: similar exciton densities ($\sim 10^{12} cm^{-2}$) were considered in~\cite{Qi2023} for even smaller binding energy for the excitons. Moreover, as will be evident later,  our results do not depend crucially on a high density of excitons. For the long-range interactions between excitons, decaying as $\sim r^{-3}$, bosons undergo a BKT transition~\cite{Filinov2005}. To simplify the situation, we consider a contact interactions, that, however, capture the linear spectrum at small momenta. The interactions determine the phase fluctuations and a condensate description of the problem is justified for scales of the sample smaller than the typical coherence length and low temperatures~\cite{Petrov2004Oct, Boudjema2013}.
 In this parameter regime we expect to our system realizes a good superfluid~\cite{Filinov2010}. Even though detailed numerical studies are lacking for the specific model, we further argue that the condensate fraction is expected to be sizeable as well~\cite{Arrigoni2011}.

\begin{figure}[t]
\includegraphics[width=0.49\textwidth]{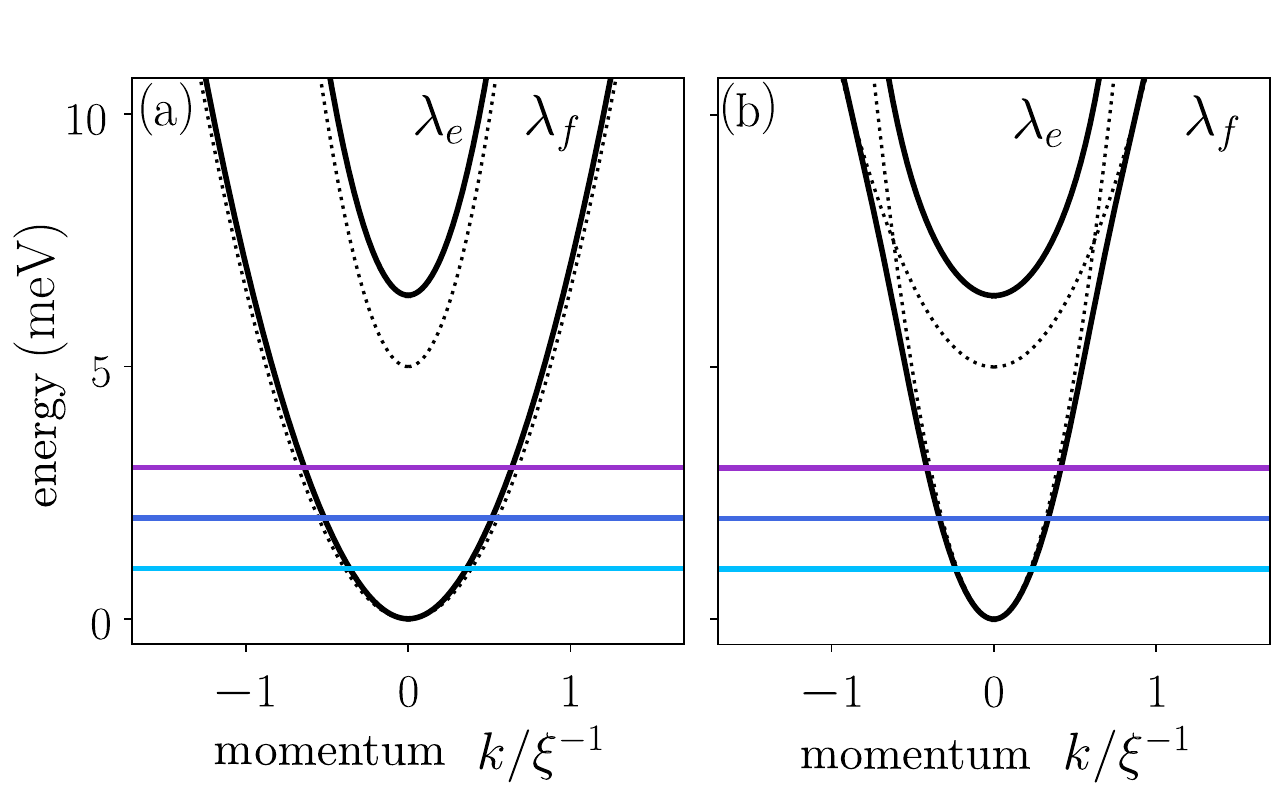}
\caption{\textbf{Hybridized trion and hole bands.} Hybridized bands $\lambda_{e/f}(k)$ (solid lines) and bare bands (dotted lines) as a function of momentum $k$ in units of the inverse exciton condensate healing length $\xi^{-1}$. The Fermi energy (horizontal lines) is chosen to be below the upper band. (a) For $\Delta<0$, the heavier band is filled, implying that doped charges mainly have trion character. (b) For $\Delta >0$, the lighter band is filled and the doped charges will mainly have hole character.  
}
\label{fig:2}
\end{figure}

\textbf{Effective attractive interactions.---}To determine the interactions between doped charges we diagonalize the free fermionic Hamiltonian~\cite{Powell05,Marchetti08}: $\hat{H}_0=\hat H_\text{holes} + \hat H_\text{trions}  +g\sqrt{n_0} \sum_{\mathbf{k}} \big[\hat{m}^\dagger_{\mathbf{k}}\hat{c}_{\mathbf{k}}+ \text{h.c.}\big]$ to obtain the dressed fermions $\hat{f}_{\mathbf{k}}=\alpha_{\mathbf{k}}\hat{m}_{\mathbf{k}}-\beta_{\mathbf{k}}\hat{c}_{\mathbf{k}}$ and $\hat{e}_{\mathbf{k}}=\alpha_{\mathbf{k}}\hat{m}_{\mathbf{k}}+\beta_{\mathbf{k}}\hat{c}_{\mathbf{k}}$, with eigenenergies 
$\lambda_{e,f}(\mathbf{k})=\frac{k^2}{4}\frac{M+m}{Mm}+\frac{\Delta}{2} \pm \sqrt{\left(\frac{k^2}{4}\frac{M-m}{Mm}-\frac{\Delta}{2}\right)^2+g^2 n_0}$, which define the upper and lower band; see Fig.~\ref{fig:2}. The gap between the two bands, $\sqrt{\Delta^2+4 g^2 n_0}$, is finite for all values of the electric field and has the minimal value of $2g \sqrt{n_0}$ for $\Delta =0$. Assuming the Fermi energy to be lower than the energy difference between the two bands and the temperature to be smaller than the gap, we restrict our analysis the lowest filled band. Then, depending on the sign of the electric field $\Delta$, charges will mainly have trion ($\Delta < 0$) or hole ($\Delta >0$) character. 
The resulting Hamiltonian of fermions $\hat f$ and condensate excitations $\hat b$ is then
$\hat{H}= \sum_{\mathbf{k}} [\lambda_{f }(\mathbf{k}) -\mu]\hat{f}_{\mathbf{k}}^\dagger\hat{f}_{\mathbf{k}} - g \sum_{\mathbf{k}\not=\mathbf{k}'}\big[\alpha_{\mathbf{k}} \beta_{\mathbf{k}'} \hat{f}^\dagger_{\mathbf{k}} \hat{f}_{\mathbf{k'}}\hat{b}_{\mathbf{k}-\mathbf{k}'} + \text{h.c.}\big]$.

We integrate out the condensate interactions using Wegner flow equations~\cite{Wegner2000}, which gives us an effective interaction between charge carries mediated by the excitons. In contrast to a Schrieffer-Wolff transformation this leads to a divergence free potential~\cite{Glazek1994,Mielke1997,Wegner2000,Cotle16}; see supplemental material~\cite{supp}. 
 The resulting effective Hamiltonian is then:
 \begin{equation}
 \hat{H}^\text{eff}=\sum_{\mathbf{k}} [\lambda_f(\mathbf{k})-\mu] \hat{f}^\dagger_{\mathbf{k}} \hat{f}_{\mathbf{k}} + \sum_{\mathbf k,\mathbf q} V_{k,q} \hat{f}^\dagger_{\mathbf{k}-\mathbf{q}} \hat{f}^\dagger_{-\mathbf{k}+\mathbf{q}} \hat{f}_{-\mathbf{k}} \hat{f}_{\mathbf{k}},
 \end{equation}
with the interactions $V_{k,q}=-\frac{M_{\mathbf{k},\mathbf{q}}  M^*_{-\mathbf{k}+\mathbf{q},\mathbf{q}} e_x(\mathbf{q})}{[\lambda_f(\mathbf{k})-\lambda_f(\mathbf{k}-\mathbf{q})]^2+e_x(\mathbf{q})^2}$ and $M_{k,q}=-g \alpha_{\mathbf{k}} \beta_{\mathbf{k}-\mathbf{q}}$. The effective interaction is thus attractive and directly proportional to the hybridization of molecule and hole bands.

\textbf{Ginzburg-Landau analysis.---}To analyze the superconducting order, we identify the pairing instability by looking for unstable solutions in the temporal evolution of the superconducting order parameter~\cite{abrikosov1975methods,Pekker2011,Knap2016}. In a mean-field formalism, the dynamics of the Cooper pair annihilation operator is
\begin{equation}\label{eq5}
  \frac{d}{dt} \langle\hat{f}_{\mathbf{k}} \hat{f}_{-\mathbf{k}}\rangle= -\frac{i}{\hbar} \sum_{\mathbf{p}>0}\big[A_{\mathbf{k},\mathbf{p}} + B_{\mathbf{k},\mathbf{p}} \big]\langle\hat{f}_{\mathbf{p}} \hat{f}_{-\mathbf{p}}\rangle,
 \end{equation} 
with contributions from the non-interacting part of the Hamiltonian $A_{\mathbf{k},\mathbf{p}}= 2 (\lambda_f(\mathbf{k})-\mu)\delta_{\mathbf{k}, \mathbf{p}}$ and the two-body interaction $B_{\mathbf{k},\mathbf{p}}=-( 2n_{\mathbf{k}}-1)\big[V_{\mathbf{k},\mathbf{k}-\mathbf{p}}-V_{\mathbf{k},\mathbf{k}+\mathbf{p}}-V_{\mathbf{-k},\mathbf{-k}-\mathbf{p}}+V_{-\mathbf{k},-\mathbf{k}+\mathbf{p}}\big]$; $n_{\mathbf{k}} =\langle \hat{f}^\dagger_{\mathbf{k}}\hat{f}_{\mathbf{k}}\rangle$ is the thermal expectation value.  Here, both $A$ and $B$ are real matrices. The physical interpretation is that if the thermal state is susceptible to form superconductivity, we expect the pairing correlations $\langle \hat{f}_\mathbf{k}\hat{f}_\mathbf{-k} \rangle $ to grow in time, indicating an unstable solution. To study this instability, we write the time evolution of  $\psi_{\mathbf k}=\langle\hat{f}_{\mathbf{k}}\hat{f}_{-\mathbf{k}}\rangle$  by means of time-dependent Ginzburg Landau equations~\cite{Stephen_tdGL_1964,Anderson_tdGL_1965,Abrahams1966} 
\begin{equation} \label{ginzburgevo} 
\frac{d}{dt}{\psi}_{\mathbf k}(t)=-\frac{\delta\mathcal{F}^{GL}}{\delta \psi^*_ {\mathbf k}}=-\sum_{{\mathbf p} > 0}F^{GL}_{\mathbf k,\mathbf p}\psi_{\mathbf p}.\end{equation}  
Here $\mathcal{F}^\text{GL}$ is the Ginzburg-Landau free energy, which to leading-order reads \begin{equation}\label{gl}
    \mathcal{F}^\text{GL}=\sum_{\mathbf k,\mathbf p} \psi_{\mathbf k}^* F^\text{GL}_{\mathbf k,\mathbf p}\psi_{\mathbf p} + \mathcal{O}(\psi^4).
\end{equation}  The associated eigenvectors then determine the structure of the unstable mode. In the Ginzburg-Landau formalism, the eigenvalues of the matrix $F^\text{GL}$ are proportional to $\propto T-T_c$. Hence, the dominant eigenvalues change their sign when driving the system into the ordered phase.  This formalism allows us to identify any possible pairing instability and then to select the most relevant one with the highest critical temperature. By contrast, a mean-field decomposition would require to select the most relevant pairing channel on before hand.
Comparing Eq.~\eqref{ginzburgevo} and Eq.~\eqref{eq5}, we obtain the relation $-i\hbar F^{GL} =C=A+B$; the set of equations can be written in a matrix form as $i\hbar\frac{d}{dt}|{\psi}\rangle= C|\psi\rangle$. The superconducting gap order parameter is related to $\psi_{\mathbf k}$ by  $\Gamma_{\mathbf{k}}= \sum_{\mathbf{p}} V_{\mathbf{k},\mathbf{p}-\mathbf{k}}\psi_{\mathbf{p}}$. 

We solve the differential equation by finding the eigenvectors $|\psi\rangle$ and eigenvalues $E_\psi$ of $-i \hbar F^\text{GL}$. The critical point is then obtained by searching for the temperature at which the imaginary part of some of the $E_\psi$ turns positive, indicating the onset of the unstable mode. From the associated eigenvector, we infer the momentum-space structure of the instability. 
By expanding the order parameter in its angular momentum contribution we find numerically that p-wave is the only relevant component; see supplemental material~\cite{supp}.

The precise p-wave channel that drives the superconducting instability can be determined separately by minimization of the free energy. 
To this end, we compare the free energy for the chiral (p$_x$+ip$_y$) and nematic (p$_x$ or p$_y$, respectively) phases. We find that for the instabilities determined as above, the free energy is indeed minimized for the topological p$_x$+ip$_y$ superconductor, see supplementary materials ~\cite{supp}, which hosts chiral Majorana edge modes. The Ginzburg-Landau analysis identifies the leading instability, and hence captures the phases near the transition temperature $T_c$. At even lower temperatures, non-linearities may stabilize other competing phases.

\begin{figure}
\centering
\includegraphics[width=0.47\textwidth]{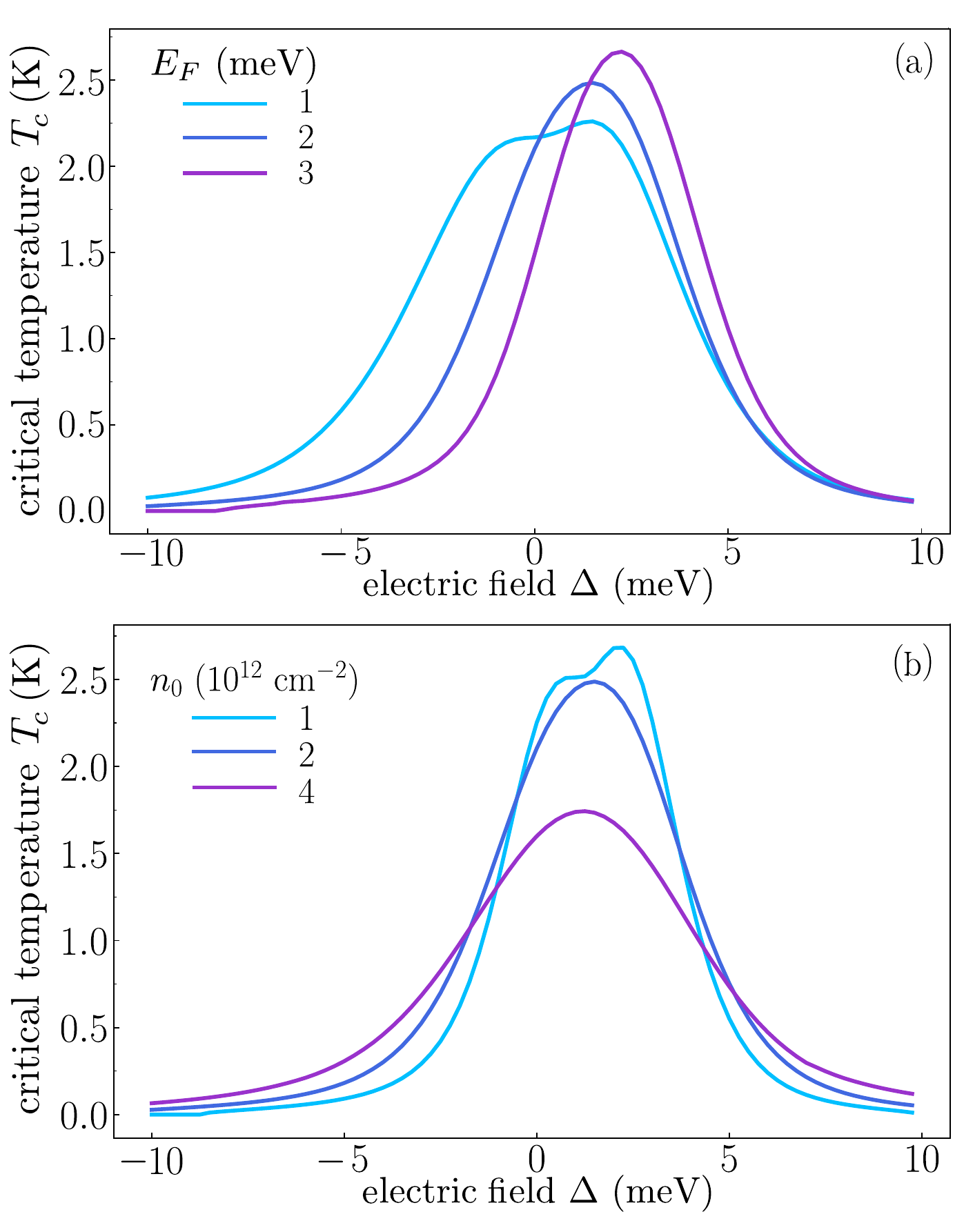}
\caption{\textbf{Critical temperature.} We evaluate the critical temperature $T_c$ as a function of the electric field $\Delta$ for (a) different Fermi energies at condensate density $n_0 = 2 $ $10^{12}$ cm${}^{-2}$ and for (b)  different condensate densities and Fermi energy $E_F = 2$ meV. By tuning the electric field $\Delta$ we find that superconductivity can be resonantly enhanced.}
\label{fig:3}
\end{figure}

\textbf{Tuning topological superconductivity.---}We have seen that the attractive interactions sensitively depend on the electric field via the hybridization of molecule and hole bands. This suggests that the superconducting order is controllable by a solid state Feshbach resonance. When adjusting $\Delta$ only over a couple of meV, we find that the superconducting transition temperature is tuned by more than an order of magnitude; see~\fig{fig:3}. The shape of the critical temperature  as a function of $\Delta$ qualitatively follows the numerically evaluated effective potential in the vicinity of the Fermi surface. By tuning $\Delta$, the hybridization between the two bands can be increased at the relevant energy scales and consequently enhance the critical temperature $T_c$. The largest $T_c$ is obtained when the hybridization of the bands is largest, which arises approximately when $\Delta \sim 3$meV.  The regime where the superconductivity is suppressed is similarly associated with a decreased hybridization of molecules and holes in proximity of the Fermi surface. 
When increasing the Fermi energy, we find from our numerical analysis that the maximum of the critical temperature  increases;~\figc{fig:3}{a}.This can be understood by the hybridization between the light hole and heavy trion band. When increasing the Fermi energy the band becomes more trion-like for $\Delta \sim |E_T|>0$. Hence, the hybridized charge carriers $\hat{f}$ attain a larger effective mass when increasing the Fermi energy. This in turn leads to a larger critical temperature.
 
Varying the condensate density $n_0$ at a fixed Fermi energy changes both the shape of the hybridized bands $\lambda_{e/f}(\mathbf k)$ and the condensate healing length $\xi=1/\sqrt{2m_x U n_0}$. The healing length determines the length scale at which the condensate effectively responds to a perturbation. For larger healing length, we thus expect that the effective attractive interactions are mediated more efficiently; \figc{fig:3}{b}. 
This is further supported by evaluating the effective attractive interaction for scattering near the Fermi surface which for small fields is given by $V_{k,q} \sim \sqrt{\frac{m_x}{U n_0 q^2}}$. The attractive potential near the Fermi surface thus increases when the condensate density decreases, as expected from the scaling of the healing length $\xi$. While being tunable by the electric field, the critical temperature depends on the thickness of the hBN, that changes the contact interaction parameter $g$ through the tunneling parameter $t$. Thus the critical temperature decreases as $g$ is decreased, see supplemental material~\cite{supp}.


\textbf{Competing effects.---}
In the case, of an imperfect but still strongly imbalanced polarization of excitons (say dominant creation of $K$ valley excitons and small density of $K'$ valley excitons), an s-wave instability could develop, provided the two exciton species interact. However, due to the strong imbalance of the Fermi surfaces of the two formed trion species, s-wave superconductivity can only occur at finite momenta and is in general energetically disfavored compared to the normal state. If it was nonetheless allowed to form, it would be strongly suppressed compared to the p-wave channel; see supplemental material~\cite{supp}.

Another competing process is the formation of a Wigner crystal in the hole-doped layer which arises at very low carrier densities~\cite{Smoleski2021}. Thus, this regime can be avoided by increasing the charge density. 

In cold atom Bose-Fermi mixtures often phase separation arises~\cite{Marchetti08,Suzuki2008}. By contrast in the set-up considered here, phase separation is disfavored by long-range Coulomb interactions between the holes, as an increased hole density in phase separated regions leads to a strong increase of the Coulomb interactions. 

\textbf{Discussion and Outlook.---}In this work, we have demonstrated the potential of two-dimensional TMD heterostructures as a tunable platform for realizing correlated Bose-Fermi mixtures of interlayer excitons and doped charges. We have shown that at low temperatures the system stabilizes p-wave topological superconductivity. We propose to utilize solid state Feshbach resonances to modify the effective interaction strength between the constituents.
Experimentally, the superconducting order parameter can be extracted by measuring the ac conductivity. In the superconducting phase the dissipative response of the conductivity then only exhibits spectral weight at frequencies above twice the gap. Within BCS theory, the critical temperature sets the energy scale for the gap and this places our estimates of these frequencies below $\sim 200$ GHz, which are in the experimentally accessible microwave regime. First experimental signatures for the superconducting phase could also be obtained from the single-particle gap, measured optically via the change of the line shape of the top layer exciton~\cite{Shimazaki2020}. 
 Other possible experimental probes may directly be able to detect the chiral edge states of a p-wave superconductor. Those include spin-polarized Scanning Tunneling Microscopy (STM)~\cite{Jack2021,Feldmaier2020}, Josephson STM~\cite{Jack2016,Randeria2016}, and the Quantum Twisting Microscope~\cite{pichler2024,Inbar2023}. These techniques could detect the Majorana zero modes as a signature of topological superconductivity and  distinguish between s-wave superconductivity and p-wave superconductivity.

For future studies it will be interesting to explore competing instabilities at low temperatures, such as nematic or density-wave phases, for which non-linearities and the competition of order parameters need to be taken into account. In this work, we consider the regime of high condensate densities, such that a possible
back-action of the charge carriers on the exciton condensate is largely negligible. New exciting phenomena appear in degenerate Bose-Fermi mixtures for low condensate densities, which are interesting to be explored both theoretically and experimentally~\cite{DeMarco2019, Duda2023, Yan2023}.

\textbf{Note added.---}Milczewski \textit{et al.} explored a complementary scheme for inducing s-wave superconductivity in Bose-Fermi mixtures realized with TMD heterostructures~\cite{Milczewski2023}. 

\textbf{Data and Code availability.---}Data analysis and simulation codes are available on Zenodo upon reasonable request~\cite{zenodo}.

\textbf{Acknowledgements.---}We thank L. Classen, E. Demler, A. Keselman, J. Knolle, and R. Schmidt, for insightful discussions. 
We acknowledge support from the Deutsche Forschungsgemeinschaft (DFG, German Research Foundation) under Germany's Excellence Strategy--EXC--2111--390814868, DFG grants No. KN1254/1-2, KN1254/2-1, and TRR 360 - 492547816 and from the European Research Council (ERC) under the European Unions Horizon 2020 research and innovation programme (Grant Agreement No. 851161), as well as the Munich Quantum Valley, which is supported by the Bavarian state government with funds from the Hightech Agenda Bayern Plus. The work of A.I. was supported by the Swiss National Science Foundation (SNSF) under Grant Number 200021-204076.
\bibliography{review/new/bibliov2} 
\section{Supplemental Material}
\subsection{Effective two-channel model}
We obtain the effective Hamiltonian of Eq. (1) in the main text starting from a microscopic model of excitons and doped charges:
\begin{equation}
\begin{aligned}
\hat{H} =& \sum_{\mathbf{k}} x^\dagger_{\mathbf{k}} \frac{k^2}{2m_x} x_{\mathbf{k}} +\begin{pmatrix}c^\dagger_{\mathbf{k},T} \\ 
c^\dagger_{{\mathbf{k}},M}\end{pmatrix} \begin{pmatrix} \frac{{k}^2}{2m}+ \Delta & t \\ t & \frac{k^2}{2m}\end{pmatrix}\begin{pmatrix} c_{{\mathbf{k}},T} \\ 
c_{{\mathbf{k}},M}\end{pmatrix}\\
&+ \frac{\tilde{U}}{V} \sum_{\mathbf{k},\mathbf{k}',\mathbf{q}} c_{{\mathbf{k}},M}^\dagger c_{{\mathbf{k}}+{\mathbf{q}},M} x^\dagger_{{\mathbf{k}'}} x_{{\mathbf{k}'}-{\mathbf{q}}},
\label{Eq:BoseFermiHamiltonian}
\end{aligned}
\end{equation}
where the index $\lbrace T,M\rbrace$ labels the top and middle layer; the interactions between inter-layer excitons and holes is modelled by an effective contact interaction of strength $\tilde{U}$. In the dilute limit, this model leads to the Hamiltonian Eq. (1) reported in the main text provided the scattering amplitudes between excitons and electrons are correctly matched. Other bound states, such as intralayer excitions or few-body complexes, are in principle present in the system as well. However, these arise at vastly different energy scales. When optically exciting the system, one can therefore select energetically which type of excition should be created~\cite{Barre2022,Wietek2024}.

Close to a Feshbach resonance, when $\Delta \simeq E_{T}^0$, the scattering is dominated by the inter-layer trion bound state. We can thus consider an effective model for the T-Matrix where scattering between exciton and top layer electrons is mediated by a virtual molecule. Assuming $g$ is the bare molecule-exciton-hole coupling, the off-shell T-matrix to second order in $g$ reads $T^{E}(\omega)\sim\frac{g^2}{\omega-E^0_T+\Delta+i\epsilon}$, which has a pole at the molecular energy. 

When considering the microscopic model for excitons and charge carriers \textit{in the same layer}, the scattering amplitude in the vicinity of the resonance is $\frac{2\pi E^0_T}{\mu_\text{red}(\omega-E^0_T+\Delta+i\epsilon)}$, where $\mu_\text{red} = m\, m_x / (m+m_x)$ is the reduced mass of the hole and the exciton. This relation is obtained by expanding the logarithm of two-dimensional T-matrix~\cite{Clemens21}. The tunneling process can be added perturbatively to this by introducing a vertex with tunneling strength $t$ and a propagator that describes tunneling of the upper layer charge carrier to the middle layer. Then, close to resonance, the tunneling process to the middle layer and back contributes a factor $\frac{t^2}{\Delta^2}$. The global T-matrix is thus $\sim \frac{2\pi E^0_T}{\mu_\text{red}\Delta^2(\omega-E^0_T+\Delta+i\epsilon)} t^2 $. Comparing the scattering amplitudes to second order in the tunneling strength $t$ yields a simple matching condition:
\begin{equation}
 g = \sqrt{\frac{2\pi}{\mu_\text{red} E_{T}^0 }}t,
\end{equation}
relating the scattering matrix in the microscopic and effective model.

\subsection{Imperfect Exciton Polarization} Here we analyze a less stringent experimental scenario and allow for the formation of a small density of excitons in the opposite polarization, due to experimental imperfections. The Hamiltonian that describes the excitons is
\begin{equation}
    H=H^{\uparrow}+ H^{\downarrow}+ \frac{U_{\uparrow\downarrow}}{A}\sum_{\bold p, \bold p' \bold q} x^{\uparrow\dagger}_{\bold p+\bold q}x^{\downarrow\dagger}_{\bold p'-\bold q}x^{\downarrow}_{\bold p'}x^{\uparrow}_{\bold p}
\end{equation}
Where we called $\downarrow$ ($\uparrow$) the excitons that couple with $\downarrow$ ($\uparrow$) holes, and thus correspond to different valley polarizations. 
We assume for simplicity equal intraspecies interactions $U$ for $\uparrow$ and $\downarrow$ excitons, and intraspecies interaction $U=U_{\uparrow\downarrow}$. This approximation is justified by the fact that interlayer exciton-exciton interactions are dominated by dipole-dipole interactions. We furthermore assume that only a small density of $\downarrow$ excitons are present, while the opposite polarization is generated at high density. Then the two condensates for the $\downarrow$ and $\uparrow$ excitons coexist and the Bogoliubov excitations above the condensates $b_{1, \bold p}$, $b_{2, \bold p}$ are a linear combination of both exciton species. In this case the total Hamiltonian has the form
\begin{align} H&=H_{trion}^{\uparrow}+H_{holes}^{\uparrow}+H_{boson}+H_{trion}^{\downarrow}+H_{holes}^{\downarrow}+ \nonumber \\&+ g\sum_{\bold k,\bold p} c^{\dagger\uparrow}_{\bold k}( a_{1,\bold p}b^\dagger_{1,\bold p}+a_{2,\bold p}b^\dagger_{2,\bold p})m^\uparrow_{\bold k+\bold p}\nonumber\\&+g\sum_{k,p} c^{\dagger\downarrow}_{\bold k}( z_{1,\bold p}b^\dagger_{1,\bold p}+\nonumber z_{2,\bold p}b^\dagger_{2,\bold p})m^\downarrow_{\bold k+\bold p}\\&+g\sqrt{n_0}\sum_{\mathbf k}c^{\dagger \uparrow}_{\mathbf{k}}m^{\uparrow}_{\mathbf{k}}+g\sqrt{n_0}\sum_{\mathbf k}c^{\dagger \downarrow}_{\mathbf{k}}m^{\downarrow}_{\mathbf{k}}+\text{h.c.}
\end{align}
and $a_{1,2}$ and $z_{1,2}$ coefficients that depend on the ratio between the two condensates and the interaction between them.
We proceed similarly to what has been done in the main text: we diagonalize the mixed bands of holes and molecules for the two spins and introduce the new fermions $f_\uparrow$ and $f_\downarrow$
\begin{align}    H&=H_0^{\uparrow}+H_0^{\downarrow}+H_{bosons}\nonumber\\&+g\sum_{k,p}\alpha^{\uparrow}_{\bold k}\beta^{\uparrow}_{\bold k+ \bold p } f^{\dagger\uparrow}_{\bold k}( a_{1,p}b^\dagger_{1,\bold p}+a_{2,p}b^\dagger_{2,\bold p})f^\uparrow_{\bold k+\bold p}\nonumber \\&+g\sum_{k,p} \alpha^{\downarrow}_{\bold k}\beta^{\downarrow}f^{\dagger\downarrow}_{\bold k}( z_{1,p}b^\dagger_{1,\bold p}+z _{2,p}b^\dagger_{2,\bold p})f^\downarrow_{\bold k+\bold p}+\text{h.c.}
\end{align}
Applying the Wegner flow procedure and keeping only zero momentum pairs we obtain both decoupled p-wave and coupled s-wave channels. 
However, the s-wave superconductivity would occur between two highly imbalanced Fermi surfaces, as the density of $\downarrow$ trions in the system is limited by the number of $\downarrow$ excitons. Strongly imbalanced Fermi surfaces, however, cannot stabilize s-wave superconductivity but rather give rise to a first-order transition to normal state, when the difference between the chemical potentials $\sim \Delta_s/\sqrt{2}$, where $\Delta_s$ is the s-wave gap. Observing that the scale of energies in general is $\Delta_s \ll E_f$ this condition should be easily satisfied in experiments and thus s-wave superconductivity cannot be stabilized~\cite{Chevy2010}. 
Furthermore, even if present, the s-wave superconductivity is expected to be highly suppressed because of the small hybridization of the hole and trion bands. The product of hybridization coefficients $\alpha ^\downarrow \beta ^\downarrow$ scale as $g \sqrt{n_0^\downarrow}$; the width of the interval where the hybridization is relevant scales as $g\sqrt{n_0^\downarrow}$, and it is centered at $\frac{k^2_f(M-m)}{Mm}$, where $k_f$ is Fermi vector for the $\downarrow$ trion. Since $k^2_f\propto n_0^\downarrow$, the s-wave superconductivity is expected to be highly suppressed in the regime of highest critical temperature for p-wave superconductivity and could thus not develop. Moreover, the p-wave channel with opposite chirality is even more suppressed. In summary, this analysis shows that an imperfect valley polarized exciton creation is not detrimental for the stabilization of p-wave superconductivity in our setting. 

\subsection{Wegner Flow}
We present the Wegner flow method that we used to trace out the condensate excitations and to obtain an effective interaction between the fermions $\hat{f}$~\cite{Wegner2000}. Another approach to determine the effective interactions is a Schrieffer-Wolff transformation, which however, leads to singular contributions. As discussed in Ref.~\cite{Cotle16} both approaches give rise  to the same interaction at the Fermi surface, where the relevant physical processes take place. The Wegner flow is advantageous though, as no singularities arise in the effective potential. Since we are interested in a microscopic model of electron-electron interactions, we use this approach in our work. 
Given the full Hamiltonian
\begin{align}
 \hat{H}&= \hat H_{f,0} + \hat H_{f,b} \\ &= \sum_{\mathbf{k}} \big[\lambda_{f }(\mathbf{k}) -\mu\big]\hat{f}_{\mathbf{k}}^\dagger\hat{f}_{\mathbf{k}} +\sum_{\mathbf{k}, \mathbf{q}} \big[ M_{\mathbf{k},\mathbf{q}}\hat{f}^\dagger_{\mathbf{k}} \hat{f}_{\mathbf{k}-\mathbf{q}}\hat{b}_{\mathbf{q}} + \text{h.c.}\big],
 \end{align}
our aim is to obtain a transformation $U(l)$ depending on the continuous parameter $l$ such that the Hamiltonian $\hat{H}(l)=U^\dagger(l)\hat{H} U(l)$ depends more on $\hat f$ only as $l$ increases. The generator of the transformation is the operator $\eta(l)$, 
 \begin{equation}\label{flow}
 \frac{d}{dl }\hat{H}(l)=\big[\eta(l), \hat{H}(l)]
 \end{equation}
 and $\big[\frac{d}{dl}U(l)\big] U^\dagger (l)=\eta(l)$. 
 Following~\cite{Wegner2000}, we consider $\eta(l)=[\hat H_{f,0}(l),\hat H_{f,x}(l)]$ as an ansatz for the continuous transformation 
 \begin{equation}\label{gen}
 \eta(l)=\sum_{\mathbf{k}, \mathbf{q}} \left[ M_{\mathbf{k},\mathbf{q}}(l) \Omega(\mathbf{k}, \mathbf{q}) \hat{f}^\dagger_{\mathbf{k}}\hat{f}_{\mathbf{k}-\mathbf{q}}\hat{b}_{\mathbf{q}}- \text{h.c.}\right] .
 \end{equation}
 where $\Omega(\mathbf{k}, \mathbf{q})=\lambda_f(\mathbf{k})-\lambda_f(\mathbf{k}-\mathbf{q})-e_x(\mathbf{q})$. The flow in Eq. (11) in the main text has a contribution from the commutator $\big[\eta(l), \hat H_{f,0}]$, which yields a three body term and defines the flow of the coefficient $M_{\mathbf{k},\mathbf{q}}(l)$ as
 \begin{equation}
 \frac{d}{dl} M_{\mathbf{k},\mathbf{q}}(l)= -\Omega(\mathbf{k}, \mathbf{q})^2 M_{\mathbf{k},\mathbf{q}}(l)
 \end{equation}
Solving this equation gives $M_{\mathbf{k},\mathbf{q}}(l)=e^{-\Omega(\mathbf{k}, \mathbf{q})^2 l}M_{\mathbf{k},\mathbf{q}}(l=0)$, and hence $M_{\mathbf{k},\mathbf{q}}(l\to \infty)=0$. For $l \to \infty$ the three-body term disappears as we required from the flow. However, the commutator $[\eta(l),\hat{H}_{f,x} ]$ introduces complicated many-body terms for the $\hat f$ particles. In particular, they generate a fermion-fermion interaction $\hat H_{f,f}=\sum_{\mathbf{k}, \mathbf{k'},\mathbf{q}}$ $V_{\mathbf{k}, \mathbf{k'},\mathbf{q}} \hat{f}_{\mathbf{k}-\mathbf{q}}^\dagger \hat{f}_{\mathbf{k'}+\mathbf{q}}^\dagger \hat{f}_{\mathbf{k'}} \hat{f}_{\mathbf{k}}$. To analyze superconducting instabilities, we focus on the terms in the potential with $\mathbf{k} = -\mathbf{k}'$, which are the most relevant for the Cooper pair formation. From Eq. (11) of the main text we then obtain 
 \begin{equation}
 \frac{d}{dl} V_{k,-k,q}= M_{\mathbf{k},\mathbf{q}}(l)M_{-\mathbf{k}+\mathbf{q},\mathbf{q}}(l)\big[ \Omega(\mathbf{k}, \mathbf{q})+\Omega(-\mathbf{k}+\mathbf{q}, \mathbf{q})\big].
 \end{equation}
 By explicitly integrating the above equation we obtain the effective interaction
 \begin{equation}
 \int_0^\infty \frac{d}{dl} V_{k,-k,q}= \frac{ -M_{\mathbf{k},\mathbf{q}}(0) M^*_{-\mathbf{k}+\mathbf{q},\mathbf{q}}(0) e_x(\mathbf{q})}{(\lambda_f(\mathbf{k})-\lambda_f(\mathbf{k}-\mathbf{q}))^2+e_x(\mathbf{q})^2}. 
 \end{equation}
The commutator of Eq. (11) in the main text gives rise to additional terms as well, including density-density terms, that are not relevant for our superconductivity analysis. Other terms are of higher order in the superfluid density, that is assumed to be small, or are of higher order in the interaction $g$ and, thus can be neglected as well~\cite{Wegner2000}.
 
\subsection{ Angular momentum expansion}
To identify the contribution of each angular momentum to the superconducting order parameter, we consider an expansion of the eigenvectors of the matrix $C$ as $\psi_{\mathbf{k}}=\sum_n \psi^c_{k, n} \cos{((2n+1) \theta_k)}+\psi^s_{k, n} \sin{((2n+1) \theta_k)}$. Using Eq. (6) of the main text we can then write a set of equations for the components $\psi^c_{k, n} $ and $\psi^s_{k, n}$
 \begin{multline}
 \frac{d\psi^c_{n,k}}{dt} =-\frac{i}{\hbar}\sum_m\int d\theta_k \cos{(2n+1)\theta_k}A_{\mathbf{k}}\big[\cos{((2m+1)\theta_k )}\\\psi^c_{m,k}+\sin{((2m+1)\theta_k)} \psi^s_{m,k}\big]+\sum_{p,m}\int d\theta_k d\theta_p \cos{((2n+1)\theta_k)}\\B_{\mathbf{k},\mathbf{p}} \big[\cos{((2m+1)\theta_p) }\psi^c_{m,p}+\sin{((2m+1)\theta_p )}\psi^s_{m,p}\big]
\end{multline}
\begin{multline}\label{eig}
 \frac{d \psi^s_{n,k}}{dt} =-\frac{i}{\hbar}\sum_m\int d\theta_k \sin{((2n+1)\theta_k)}A_{\mathbf{k}}\big[\cos{((2m+1)\theta_k )} \\\psi^c_{m,k}+\sin{((2m+1)\theta_k)} \psi^s_{m,k}\big]+\sum_{p,m}\int d\theta_k d\theta_p \sin{((2n+1)\theta_k)}\\B_{\mathbf{k},\mathbf{p}} \big[\cos{((2m+1)\theta_p) }\psi^c_{m,p}+\sin{((2m+1)\theta_p )}\psi^s_{m,p}\big]
\end{multline}
Consequently, we define a projected matrix $\tilde{C}=\tilde{A}+ \tilde{B}$. The instability can then be identified by exponentially growing solutions of the order parameter as a function of time as discussed in the main text. In this new definition, $\tilde{A}$ and $\tilde{B}$ are tensors defined by indices: $\tilde{A}_{k, n,p,m}^{c,c}=\psi_{k,p}\int d\theta_k \cos{((2n+1)\theta_k)}A_{\mathbf{k}}\cos{((2m+1)\theta_k )}$, $\tilde{A}_{k, n,p,m}^{c,s}=\psi_{k,p}\int d \theta_k \cos{((2n+1)\theta_k)}A_{\mathbf{k}}\sin{((2m+1)\theta_k )}$ and $\tilde{A}_{k, n,p,m}^{s,s}$ and $\tilde{A}_{k, n,p,m}^{s,c}$ are given correspondingly by considering sine instead of cosine in $\tilde{A}_{k, n,p,m}^{c,c}$ and $\tilde{A}_{k, n,p,m}^{c,s}$. Similarly, $\tilde{B}^{c,c}_{k,n, p, m}= \int d \theta_k d \theta_p \cos{((2n+1)\theta_k)}B_{\mathbf{k},\mathbf{p}} \cos{((2m+1)\theta_p) } $, $\tilde{B}^{c,s}_{k,n, p, m}= \int d \theta_k d \theta_p \cos{((2n+1)\theta_k)}B_{\mathbf{k},\mathbf{p}} \sin{((2m+1)\theta_p) } $, and $\tilde{B}_{k, n,p,m}^{s,s}$ and $\tilde{B}_{k, n,p,m}^{s,c}$ are given correspondingly by considering sine instead of cosine and vice versa in $\tilde{B}_{k, n,p,m}^{c,c}$ and $\tilde{B}_{k, n,p,m}^{c,s}$. \\
We always find a two-fold degeneracy of the eigenvalues of the matrix $\tilde{C}$. We interpret this as due to an equal-weight contribution of the sine and cosine to the eigenvector $|\psi\rangle$. In the degenerate 2-dimensional eigenspace, the basis that diagonalizes $\tilde{C}$ can be chosen in such a way that we have an eigenvector with only sine contributions and an eigenvector with only cosine contributions. This is expected, since the Hamiltonian is symmetric under rotations. This result implies that critical temperature for the sine and cosine eigenvalue is the same.
 \subsection{Topological superconductivity} 
 \begin{figure}
 \centering
 \includegraphics[width=0.47\textwidth]{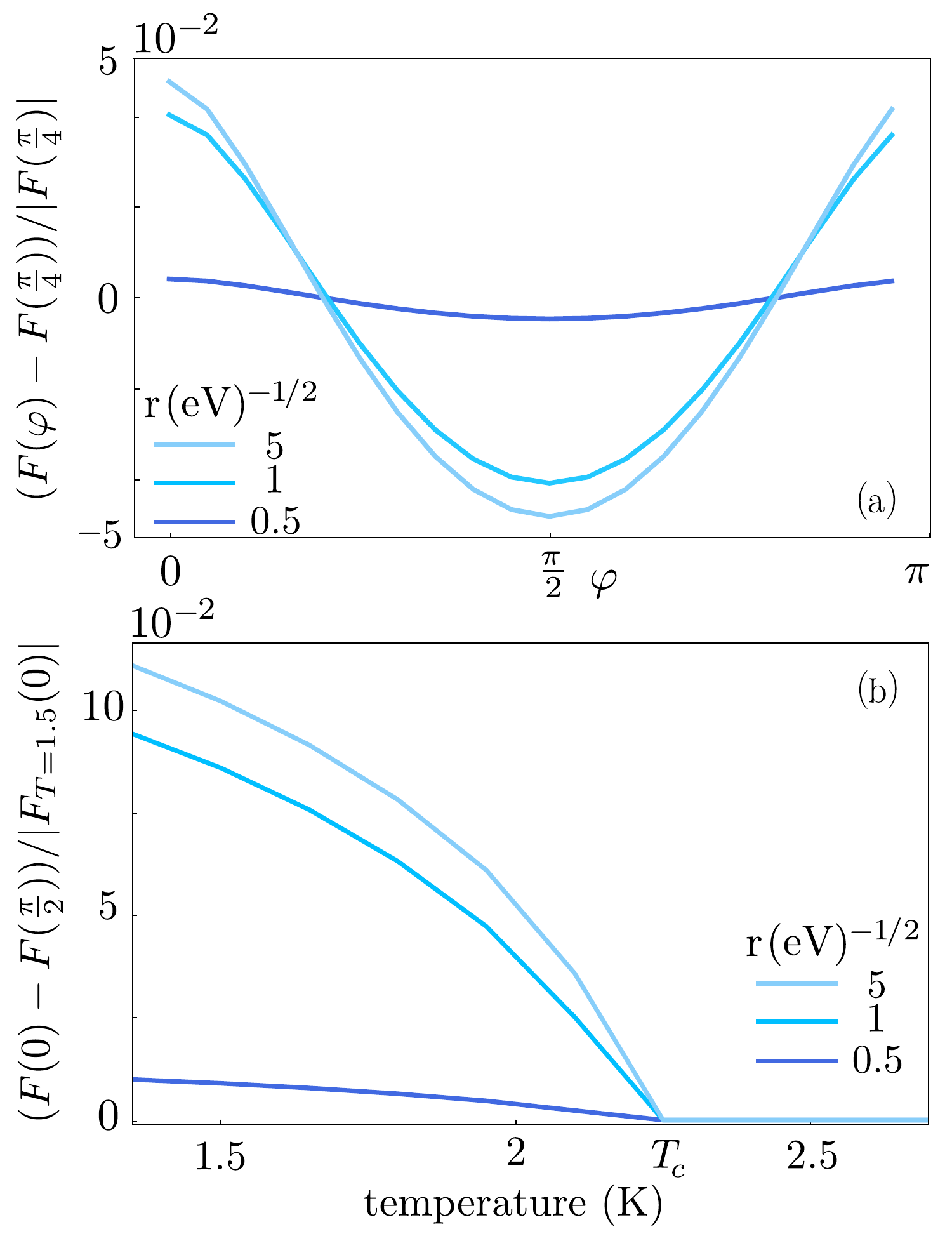}
 \caption{\textbf{Analysis of the relative phase  $\varphi$ of the angular momentum contributions to the order parameter.} (a) Free energy as a function of the phase $\varphi$, computed for $T=1.5 K$, $\Delta=2$ meV, $E_F=1$ meV. Different magnitude of the instabilities are compared by changing the parameter r, $max_k (\psi_k$)=r$\sqrt{\lambda}$, where $\lambda \propto $ T${}_c$-T and $\varrho\propto\frac{1}{r^2}$ is the fourth order term in the Ginzburg-Landau formalism, see text. (b) Difference of Free energy at $\varphi=0$ and $\varphi=\frac{\pi}{2}$ as a function of the temperature for different magnitudes of the order parameter shows that $\varphi = \frac{\pi}{2}$ is the solution with lowest free energy.}
 \label{fig:4}
 \end{figure}
Here, we discuss the symmetry of the order parameter, which leads to the conclusion that the state is a topological $p$+$ip$ superconductor.

We observed that the only relevant contribution to the eigenvector  of the maximally unstable solution $|\psi\rangle$ is given by the p-wave contribution. The instability is generated by two eigenvectors of the matrix $\tilde{C}$: one with contribution only in the cosine degrees of freedom and the other one only in the sine degrees of freedom. Furthermore, by Ginzburg-Landau analysis, we observe that the critical temperature is the same for the two directions. Because of this, the Ginzburg- Landau free energy in Eq. (4) of the main text can be written as $\mathcal{F}^{GL}= \sum \psi_x^*\lambda\psi_x+\psi_y^*\lambda\psi_y$, where we defined $\psi_x$ as the eigenstate with only cosine contributes, $\psi_y$ the eigenstate with only sine contribution, or equivalently by considering any unitary transformation of  $(\psi_x, \psi_y)$. Among these possible linear combinations, we find also the topological p-wave superconductor $(\psi_x+ i\psi_y, \psi_x -i \psi_y)$ and the nematic one $(\psi_x+\psi_y,\psi_x-\psi_y)$; up to a rotation in momentum space, the generic form of the instability can be written as $(\psi_x+ e^{i\varphi}\psi_y, \psi_x -e^{-i\varphi}\psi_y)$. To determine the nature of the p-wave superconductivity, we would have to consider further order terms in the Ginzburg-Landau free energy and find $\varphi$ such that the Ginzburg-Landau free energy is minimized. The higher-order terms can be computed perturbatively. However, in order to estimate the type of instability a simpler strategy is sufficient. We determine the order parameter up to a constant. To this end, we introduce a mean-field Hamiltonian that describes a system, that is unstable under the previous determined superconducting order parameter, and for which the only free parameter is the phase $\varphi$. By considering several different magnitudes of the order parameter, we show that the instability directions are the only relevant information needed to determine the topological nature of the superconductor, see Fig.~\ref{fig:4}. 

\begin{figure}
    \centering
    \includegraphics[width=0.5\textwidth]{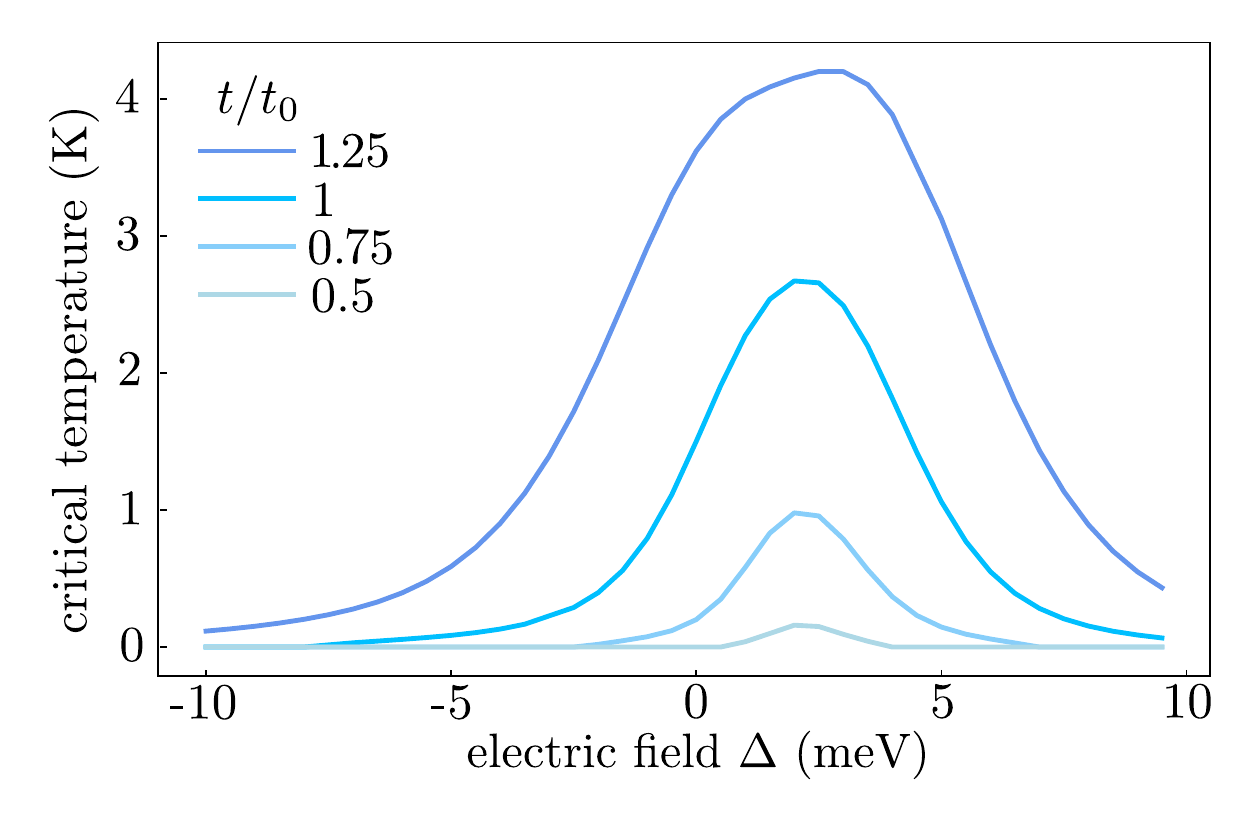}
    \caption{ { \textbf{Role of hBN layers on critical temperature.} Dependence of the critical temperature on the tunneling matrix element $t$, compared to $t_0= 0.9$meV used in the main text. The tunneling matrix element $t$ is determined by the thickness of the hBN layer.}}
    \label{fig:5}
\end{figure}

First, we show that if the only relevant contribution to the instability is p-wave, the same holds for the order parameter. Indeed we have $\Gamma_{\mathbf{k}}=\sum_{\mathbf{p}}V_{\mathbf{k},\mathbf{k}-\mathbf{p}} \psi_{\mathbf{p}}=\sum_{\mathbf{p}>0}V_{\mathbf{k},\mathbf{k}-\mathbf{p}} \psi_{\mathbf{p}}-V_{\mathbf{k},\mathbf{k}+\mathbf{p}} \psi_{\mathbf{p}}$. Using the fact that the potential is invariant under the transformation $V_{\mathbf{k}, \mathbf{q}}\to V_{-\mathbf{k}, -\mathbf{q}}$, we can write $\Gamma_{\mathbf{k}}=\frac{1}{2}\sum_{\mathbf{p}>0}V_{\mathbf{k},\mathbf{k}-\mathbf{p}} \psi_{\mathbf{p}}-V_{\mathbf{k},\mathbf{k}+\mathbf{p}} \psi_{\mathbf{p}}+V_{\mathbf{-k},\mathbf{-k}+\mathbf{p}} \psi_{\mathbf{p}}-V_{\mathbf{-k},\mathbf{-k}-\mathbf{p}} \psi_{\mathbf{p}}$. When $\psi_{\mathbf{k}}$ is an eigenvector of the matrix $C$, then 
we have from Eq~\eqref{eig} 
\begin{multline}
 2(\lambda_f(\mathbf{k})-\mu)\psi_{\mathbf k} - (2n_{\mathbf{k}}-1)\sum_{\mathbf{p}>0}\big[V_{\mathbf{k},\mathbf{k}-\mathbf{p}}+\\-V_{\mathbf{k},\mathbf{k}+\mathbf{p}}-V_{\mathbf{-k},\mathbf{-k}-\mathbf{p}}+V_{\mathbf{-k},\mathbf{-k}+\mathbf{p}}\big]\psi_{\mathbf{p}}= E_\psi  \psi_{\mathbf{k}}\end{multline}
Hence,
\begin{equation}
 2(\lambda_f(\mathbf{k})-\mu)\psi_{\mathbf k}- 2(2n_{\mathbf{k}}-1)\Gamma_{\mathbf{k}}= E_\psi \psi_{\mathbf{k}}.
 \end{equation}
from which we obtain an expression for $\Gamma_{\mathbf{k}}$
 \begin{equation}
 \Gamma_{\mathbf{k}}= \frac{2(\lambda_f(\mathbf{k})-\mu)- E_\psi}{2(2 n_{\mathbf{k}}-1)} \psi_{\mathbf{k}}
 \end{equation}
 where the only dependence from $\theta_{k}$ is coming from $\psi_{\mathbf{k}}$. This implies that to the eigenvector of the maximal unstable solution $\ket{\psi^c_{k}}$, which contributes only to the cosine degrees of freedom, corresponds to an
 order parameter, which also contributes only on the cosine degrees of freedom. The same is true for the corresponding sine degrees of freedom. Notice that $\Gamma_\mathbf{k}$ has been determined up to a constant which is the same for all momenta $\mathbf{k}$.
 
To determine whether the p-wave superconducting phase is topological or trivial, we discard the contributes of the potential that are not relevant for the formation of superconductivity and consider the mean-field model
\begin{multline}
 H= \sum_{\mathbf{k}}\xi(\mathbf{k})\hat{f}^\dagger_{\mathbf{k}}\hat{f}_{\mathbf{k}}+ \sum_{\mathbf{k}} \bigg(\big[\Gamma_k \cos(\theta_k) +\\+e^{i\varphi}\Gamma_k \sin{\theta_k} \big]\hat{f}^\dagger_{\mathbf{k}}\hat{f}^\dagger_{-\mathbf{k}} + \text{h.c.} \bigg),
\end{multline} where we defined $\xi(\mathbf{k})=\lambda_f(\mathbf{k})-\mu$. When tuning the phase $\varphi$ different p-wave superconductors can be realized, including the topological chiral and the non-topological nematic ones. Specifically, phases of $\varphi=0, \pi$ correspond to nematic order in which the only contribution to the superconductor is $p_x$ or $p_y$ and the gap is not rotational invariant leading to gapless points on the Fermi surface. By contrast, phases of $\varphi=\frac{\pi}{2}, \frac{3 \pi}{2}$ correspond to chiral topological phases in which the gap is rotational invariant and the order parameter is fully gapped. 

For this model and finite temperatures we can evaluate the free energy, $F=E-TS$; we compute the entropy as $S=-k_B [n_\mathbf{k} \ln{n_\mathbf{k}} + (1-n_\mathbf{k})\ln({1-n_\mathbf{k}})] $ using the occupation $n_\mathbf{k}$  of the Bogoliubov excitations, $n_\mathbf{k}=\frac{1}{\exp[\beta E_\mathbf{k}]+1}$, $\beta=k_B T$, $E_\mathbf{k}=\sqrt{{\xi(\mathbf{k})^2+|\Gamma_k|^2(1+\cos(\varphi)\sin(2\theta_k))}}$, and $E=\sum_{\mathbf{k}} E_{\mathbf{k}} n_{\mathbf{k}}+ E_{GS}$. We observe that the dependence on $k$ of the order parameter guarantees that $\varphi=\pm \frac{\pi}{2}$ are minima, while $\varphi= 0, \pi$ are maxima., see Fig.~\ref{fig:4} (a). We test this result against different magnitudes of the instability, effectively exploring two order of magnitudes of values for the prefactor of the fourth order term in the Ginzburg-Landau theory $\varrho \propto\frac{1}{r^2}$, defined by $\mathcal{F}=\lambda |\psi|^2 + \frac{1}{2}\varrho|\psi|^4$. In Fig.~\ref{fig:4} (b) we report the difference between the free energy at $\varphi=0$ and $\varphi=\frac{\pi}{2}$ for different values of the temperature $T\sim T_c$, showing that the same holds in a finite range of temperature in the vicinity of $T_c$. 
\subsection{Thickness of the hBN layer}

The thickness of the hBN layer between upper and middle layer determines the effective tunneling parameter $t$, which in turn changes the contact interaction parameter. We report the dependence on the critical temperature for different values of $t$, compared to the one used in the main text, see Fig.~\ref{fig:5}.

Decreasing the contact interaction parameter, decreases the strength of the effective interaction between the fermions. As a consequence the critical temperature is decreased. Furthermore, the interval of the electric field that gives superconductivity becomes smaller as the interaction is decreased. This effect is due to a decreased hybridization between the two bands. 
\end{document}